\documentclass{aa}
\usepackage{graphicx}
\begin{document}
   \title{Astronomical Seeing from the Summits of the Antarctic
   Plateau\thanks{Rodney Marks died tragically at the South Pole in
   May 2000.  This paper presents the most significant previously
   unpublished results from his PhD thesis: `Antarctic site testing:
   measurement of optical seeing at the South Pole' (Marks 2001).}}

   \subtitle{}

   \author{R.D. Marks\inst{1}}

   \offprints{Dr.\ M.G. Burton, School of Physics, UNSW, Sydney, NSW
   2052, Australia. M.Burton@unsw.edu.au}

   \institute{School of Physics, University of New South Wales, Sydney, 
              NSW 2052, Australia}

   \date{Received 12 November 2001; Revised 24 January, 2002; Accepted
   24 January 2002}

   \abstract{From the South Pole, microthermal turbulence within a
   narrow surface boundary layer some 200\,m thick provides the
   dominant contribution to the astronomical seeing.  We present
   results for the seeing at a wavelength of 2.4$\mu$m.  The narrow
   turbulence layer above the site, confined close to the surface,
   provides greatly superior conditions for adaptive optics correction
   than do temperate latitude sites.  An analysis of the available
   meteorological data for the Antarctic plateau suggests that sites
   on its summit, such as Domes A and C, probably experience
   significantly better boundary layer seeing than does the South
   Pole.  In addition, the inversion layers may be significantly
   narrower, lending the sites even further to adaptive optics
   correction than does the Pole.  \keywords{Atmospheric effects --
   Site Testing -- Antarctica -- Methods: observational} }

   \maketitle

%
%
\section{Introduction}
The effect of atmospheric turbulence on astronomical image quality, or
``seeing'', has been studied at the South Pole through measurement of
the microthermal fluctuations associated with the turbulence (Marks et
al.\ 1996---Paper I, Marks et al.\ 1999---Paper II).  The seeing was
found to be dominated by the contributions from a narrow but turbulent
boundary layer, with a minimal contribution from the free atmosphere
above it.  This suggests that, if the effects of the surface boundary
layer over the Antarctic plateau can be mitigated, superb seeing
conditions might be obtained.  Since the boundary layer seeing is
strongly influenced by the inversion wind, it is possible that
exceptionally good seeing may occur at the surface from other
locations on the plateau away from the South Pole, where this wind is
reduced. Such sites may be the high points of the Antarctic
plateau---Domes A and C---where the inversion wind is almost
non-existent. Little direct evidence is available on seeing conditions
for these sites. Some clues, however, are available from
meteorological records. We examine these in this paper, and discuss
the implications for future astronomical observatories in Antarctica.

While the South Pole is convenient location, in the sense that it is
populated all year-round, it is generally agreed that there are other
sites where the observing conditions are potentially much better, in
terms of atmospheric transmission, humidity and weather, in addition
to the seeing. The South Pole lies some 1,000\,km away from Dome~A,
the highest point on the plateau at about 4,200\,m. Much of the
boundary layer turbulence at the Pole is associated with the inversion
winds which consist of cold air rolling gradually down the slope from
the highest regions of the plateau, picking up speed as they go,
before finally turning into the violent katabatic winds that are such
a well-known feature of the weather along the Antarctic coastline. It
is quite possible that the boundary layer seeing at Dome A is much
lower than at the South Pole since, although the temperature inversion
is still present, the calm winds close to the surface mean that the
mechanical mixing of the different temperature layers is minimised.

Since Dome A still remains almost totally inaccessible it is fortunate
that there are other sites for which similar comments apply. Potential
candidates include Vostok and Dome~C\@. Dome~C has the further advantage
that the infrastructure will soon be in place at this site to support
a large observatory, with construction of the year-round
Franco-Italian {\sl Concordia Station}, expected to be completed by
2004.  Obviously, direct measurements need to be made at these sites;
however the sparse information available at this stage does allow some
general comments to be made about the likely seeing conditions at the
higher plateau sites, given the results from the South Pole.

\section{Super Seeing on the Antarctic Plateau?}
The upper-atmosphere jet streams that are common in temperate
locations are associated with the upper boundary of the troposphere,
at an altitude of around 10--12\,km.  The winds can be very strong,
averaging over 40\,m\,s$^{-1}$ in some places (e.g.\ McIlveen 1992).  In
contrast, such high-altitude winds are very much weaker in the polar
regions, and the tropopause is relatively low (7--8\,km) and less
marked. In such conditions, the free atmosphere is expected to be very
stable, with smooth wind and temperature gradients leading to
exceptionally good seeing.  Since the upper atmosphere turbulence is a
significant component of the seeing at many locations, the lack of
indicators for such turbulence on the Antarctic plateau have led to
the coining of the phrase ``Super Seeing'' (Gillingham 1993) to
describe conditions that may well be superior to any other region on
earth.

Mitigating against good seeing from the surface, however, is an
intense temperature inversion, pervasive over the plateau during the
winter months, and often exceeding 0.1$^\circ$C~m$^{-1}$ averaged over
the entire boundary layer. The limited data available on boundary
layer turbulence on the plateau (Neff 1981) indicates that some very
intense optical turbulence, concentrated in quite narrow layers,
extends throughout the lower parts of the inversion layer. It has a
vertical extent of some 300--500\,m.

On the other hand, the fact that this atmospheric disturbance is
concentrated so close to the surface, in comparison to the turbulence
at other sites, might well have its own advantages. Most large
optical/IR telescopes employ some form of real-time image correction
technique, which broadly come under the title of ``adaptive
optics''. The effectiveness of such methods is severely restricted by
angular and temporal limitations set by the nature of the atmospheric
turbulence. In general, the scale of these parameters is inversely
proportional to the altitude of the turbulent layers, and hence
high-altitude turbulence is much more difficult to correct than
low-altitude disturbances (see, for example, Cowie \& Songaila 1988,
Olivier 1993). Hence, even if the boundary-layer seeing is poor, it
may be relatively easy to eliminate this component over larger areas
than is possible at other sites. A similar altitude dependence also
applies to the scintillation of stellar sources caused by atmospheric
turbulence. The vertical structure of the atmospheric turbulence over
the Antarctica plateau should therefore lend itself to more accurate
photometry, which is especially important in astroseismology and the
study of variable stars, including planetary occultations. Another
field where the Antarctic plateau offers outstanding opportunities is
that of astrometric interferometry (e.g.\ see Lloyd, Oppenheimer and
Graham, 2002). This is because the measurement error in the spatial
position of a source made using this technique is proportional to
$h^2$, where $h$ is the height above the telescope where microthermal
fluctuations occur.  On the Antarctic plateau, where these are
confined to the surface boundary layer, this factor is much smaller
than at temperate-latitude sites, where they arise from the
high-altitude jet stream.

\section{South Pole Seeing Conditions}
\label{sec:spole}
We undertook a site-testing campaign during nighttime (i.e.\ in
winter) at the South Pole over two years; during the first, the
contribution of the lower boundary layer was examined (Marks et al.\
1996---Paper I) and in the second year this was extended to 15\,km
altitude (Marks et al.\ 1999---Paper II).  The atmosphere was found to
show a marked division into two characteristic regions: (i) a highly
turbulent boundary layer (0--220\,m), with a strong temperature
inversion and wind shear, and (ii) a very stable free atmosphere.  The
mean visual seeing over 15 balloon flights in 1997 was measured to be
$1.86''$, of which the free atmosphere component was only
$0.37''$. Direct measurements of the seeing with a differential
image-motion monitor (DIMM) by Loewenstein et al.\ (1998) are
consistent with these values.

The $\sim 220$\,m height of the boundary layer at the South Pole, in
terms of the micro-turbulence, is much lower than the 300--500~m
temperature inversion that usually is regarded as defining the
boundary layer depth. This is due to the fact that the optical
turbulence intensity depends strongly on the vertical gradients of the
wind velocity and temperature inversion. Near the top of the boundary
layer, the inversion begins to flatten out at around 200--250~m, at
which point the microthermal turbulence becomes quiet, even though the
inversion continues weakly for a further 100~m or more.

In contrast to the severe optical turbulence present in the boundary
layer, the free atmosphere is very quiescent in comparison with other
sites. The inversion that often occurs in the tropopause at
mid-latitudes does not exist in Antarctica. Instead, the temperature
profile generally levels off weakly and smoothly.

These are quite different characteristics to those at temperate
latitude sites. The Chilean observatory sites, in particular, tend to
have a high proportion of the seeing caused by turbulence in a
boundary layer extending up to about 1,000--2,000~m, with a relatively
quiescent tropopause. This is in marked contrast to Mauna Kea, which
at 4,200~m is high enough to avoid boundary layer effects almost
completely (Bely 1987). Here, however, an upper level jet stream
greatly increases the high-altitude component of the seeing, so that,
overall, the site quality is no better than the high desert sites in
Chile.

\subsection{Infrared Seeing at the South Pole}
In Paper II we calculated the seeing profiles at optical wavelengths,
$\lambda = 5000$\,\AA\@.  Here, we show the values in the near--IR,
$\lambda = 2.4 \ \mu$m, a wavelength of particular interest because of
the particularly low value for the sky background there in Antarctica
(e.g. Ashley et al.\ 1995).  These profiles are calculated assuming
that the Fried parameter, $r_0$, scales as $\lambda^{6/5}$ and the
seeing, $\varepsilon$, as $\lambda^{-1/5}$ (see Paper II).

\begin{table*}
\caption{Integrated seeing and boundary layer contributions at 
2.4$\mu$m for the South Pole, determined from 15 balloon launches
between 20 June and 18 August 1995.  The ``free atmosphere'' refers to
the entire atmosphere, excluding the boundary layer.}
\label{ir_tab}
\begin{center}
\begin{tabular}{lllllll}
\hline\noalign{\smallskip}
Measurement & Mean & Std.~Dev. & Median & Best 25\% & Best & Worst \\
\noalign{\smallskip}
\hline\noalign{\smallskip}
Seeing (arcseconds) \\
\noalign{\smallskip}
\em{~--total} & 1.36 & 0.55 & 1.2 & 0.7 & 0.6 & 2.3 \\
\noalign{\smallskip}
\em{~--free atmosphere} & 0.27 & 0.05 & 0.23 & 0.21 & 0.17 & 0.38 \\
\noalign{\smallskip}
$r_0$ (cm) \\
\noalign{\smallskip}
{\em~--total} & 36 & 22 & 42 & 65 & 81 & 22 \\
\noalign{\smallskip}
\em{~--free atmosphere} & 179 & 47 & 185 & 228 & 293 &  153 \\
\noalign{\smallskip}
\hline
\end{tabular}
\end{center}
\end{table*}

\begin{figure*}
\vskip 3cm
\resizebox{\hsize}{!}{\includegraphics{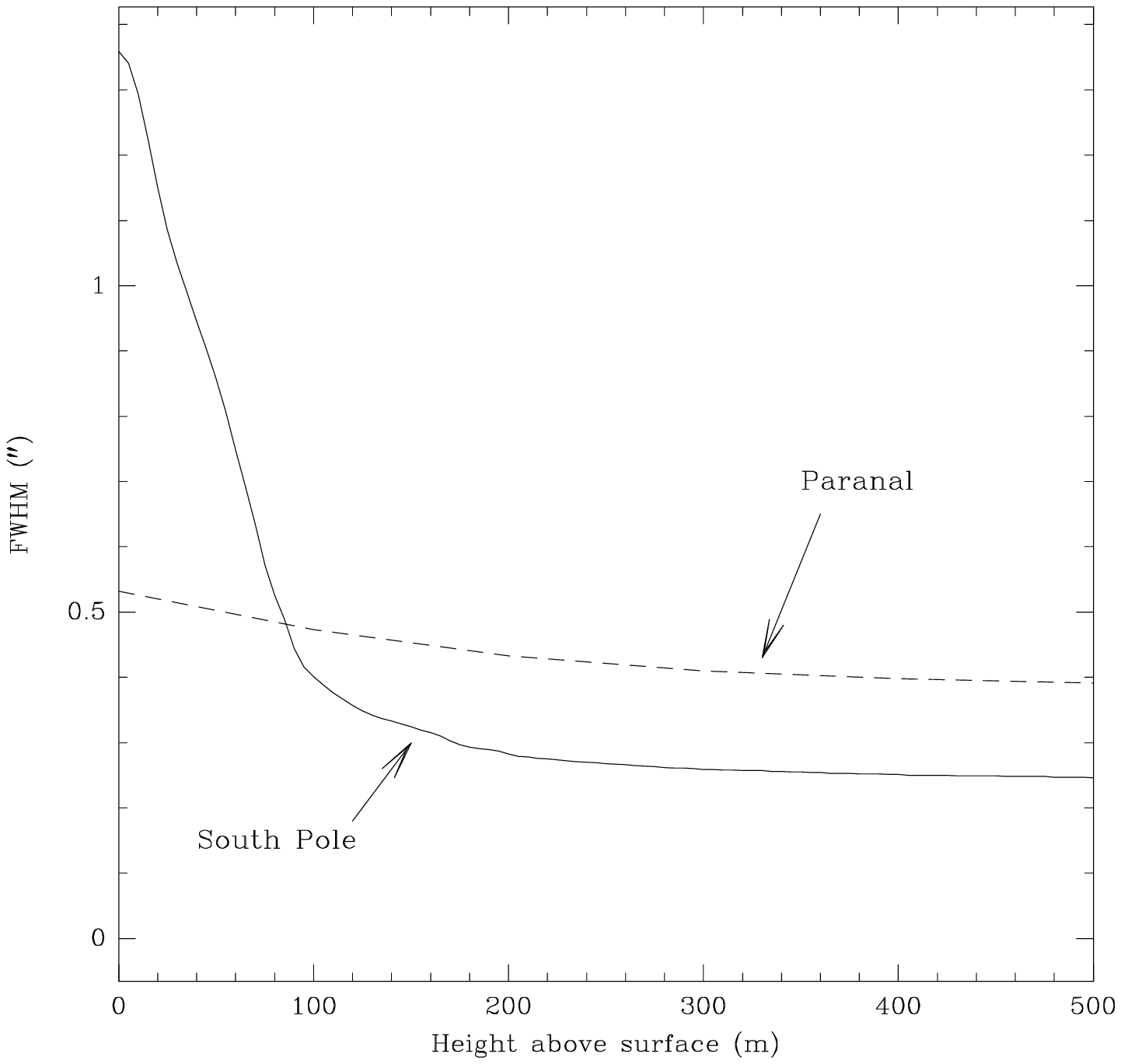}}
\caption{Seeing as a function of height of telescope above the surface,
compared between the South Pole and Cerro Paranal in Chile, for a
wavelength of $2.4\mu$m. The solid line represents the results from
balloon launches at the South Pole (Paper II), and the dashed line a
summary of a similar experiment performed at the ESO--VLT site in
Paranal (Fuchs 1995).}
\label{sp_par_ir}
\end{figure*}

The FWHM seeing and $r_0$ values at 2.4$\mu$m are shown in
Table~\ref{ir_tab}. Fig.~\ref{sp_par_ir} shows a comparison between
the seeing profiles at Paranal and the South Pole at 2.4$\mu$m. Every
tenth of an arcsecond gain is increasingly significant at IR
wavelengths, as the seeing reaches very low levels. The free
atmosphere seeing at the South Pole approaches 0.2$''$, while the
Paranal curve remains at around 0.4$''$. While the free atmosphere
seeing at the South Pole is still about 65\% of that at Paranal from
the same level, in terms of the actual values the difference between
the sites is probably more notable here than in the visible.

\begin{figure*}
\vskip -1cm
\resizebox{\hsize}{!}{\includegraphics{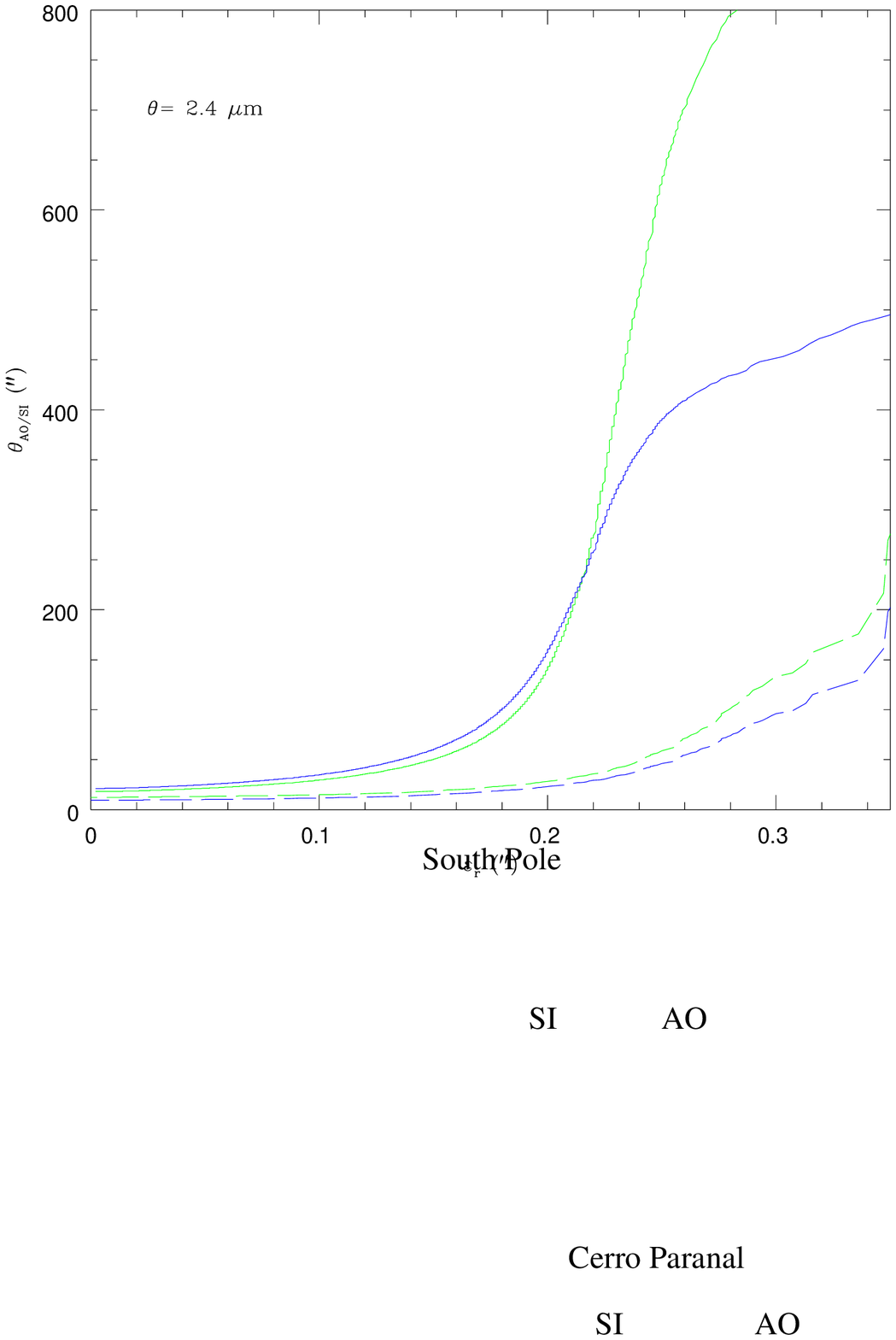}}
\caption{Isoplanatic angles, $\theta_{\rm AO}$ and $\theta_{\rm SI}$, 
as a function of the residual seeing, $\varepsilon_r$, in arcseconds
for $\lambda = 2.4\mu$m. Solid lines are derived from the average
$C_{\rm N}^2$ profile at the South Pole, and dashed lines are from
Cerro Paranal data.  The dark lines refer to adaptive optics (AO)
correction, and the light lines to speckle interferometry (SI).}
\label{ao_ir}
\end{figure*}

The adaptive optics parameters also register a corresponding
improvement, scaling as they do with $r_0$.  Assuming a Kolmogorov
spectrum, these values for optical wavelengths (Paper II) are
increased by a factor of $(2.4/0.5)^{6/5}=6.6$ when observing at
2.4$\mu$m. Hence, the isoplanatic angles, $\theta_{\rm AO,SI}$ (AO, SI
= adaptive optics, speckle interferometry---see Paper II), over the
entire atmosphere, are around 18--20$''$, while coherence times
$\tau_{\rm AO,SI}$ increase to 10--100\,ms. Angles obtained if
boundary layer correction is applied increase to very significant
values of around 7--12$'$. This is large enough to give close to 100\%
sky coverage for finding suitable stars for use in wavefront
correction, up to at least $m_{\rm K}=10$. The results are shown in
Fig.~\ref{ao_ir}.

\subsection{Forecasting the Seeing}
\label{sec:forecast}
A striking feature of our data (Paper II) was that the boundary layer
turbulence structure is not evenly spread throughout the inversion
layer but, rather, is concentrated in anywhere from one to half a
dozen narrow strips, generally 10--20~m in depth. They are almost
always associated with a sharp peak in either the potential
temperature or wind velocity gradient, or both. Quite often, the
strongest turbulence above the surface layer appears close to the top
of the boundary layer ($\sim$200\,m).

It would appear that the Richardson number, $R_i =
\frac{g}{\theta}\frac{\left({\rm d}\theta/{\rm d}z\right)}{\left({\rm
d}\vec{U}/{\rm d}z\right)^{2}}$, is a reasonable indicator of
upper-air turbulence in the atmosphere at the South Pole (see Paper
II). Here ${\rm d}\theta/{\rm d}z$ is the potential temperature
gradient, with $\theta(z)=T(z)
\left(\frac{P(z)}{1000}\right)^{-0.286}$, such that the value of $\theta$
corresponds to the temperature of the air adjusted adiabatically to a
standard pressure of 1000\,hPa.  ${\rm d}U/{\rm d}z$ is the wind
velocity gradient, $T(z)$ the temperature, $P(z)$ the air pressure,
$g$ the acceleration due to gravity and $z$ the height above sea
level.  When $R_i < \frac{1}{4}$, relatively strong upper-air
turbulence can be expected. This suggests that it may be possible to
determine the frequency of very good free atmosphere seeing at the
site by analysis of historic and ongoing meteorological records. The
importance of this is that if, in practice, it is possible to correct
for the boundary layer turbulence by the use of adaptive optics, then
the free atmosphere component of the seeing will be the limiting
factor in large telescope image resolution at the site. One could then
hope to forecast, or at least ``nowcast'' (Murtagh \& Sarazin 1993,
1995), seeing conditions in terms of just a few parameters easily
obtainable from standard balloon launches performed daily by the
Meteorology department. The infrequent formation of the shear layers
described above could perhaps be reported some time in advance,
depending on their typical lifetime, and then be used to generate an
approximate value for the upper atmosphere seeing at that time.

\section{Implications for Adaptive Optics Wavefront Correction}
The turbulence structure of the atmosphere also has implications for
the feasibility of image correction techniques such as adaptive optics
and speckle interferometry. In particular, the isoplanatic angle and
wavefront coherence time --- the main spatial and temporal parameters
limiting any image correction system --- are highly dependent on the
altitude of the strongest turbulent layers, and can be calculated from
the known $C_{\rm N}^{2}$ profiles at a given site (e.g.\ Vernin \&
Mu\~non-Tu\~n\'oz 1994, Fuchs 1995). A turbulence profile showing a
higher proportion of boundary layer turbulence, with a relatively
clear free atmosphere, will be much more tractable for a low-order
adaptive optics system than one with, for example, strong jet
stream-related turbulence in the tropopause. Most of the new
generation of large telescopes are utilising some form of image
correction of this type, and so the vertical profile becomes even more
important with these considerations in mind.

The limiting factors in any adaptive optics method are the spatial and
temporal coherence of the wavefront that one is trying to correct. In
general, the high-altitude turbulence that forms a large proportion of
the seeing at many mid-latitude sites is very difficult to correct,
and sophisticated methods are usually required to extract much
improvement in image quality (e.g.\ Cowie \& Songaila 1988), since it
is isoplanatic over an angular scale of only a few arcseconds.
However, over the Antarctic plateau we have the exact opposite
scenario, where a large proportion of the turbulence is produced very
close to the aperture of the telescope, and so the isoplanatic angle
should be much larger. Indeed, considering the single large turbulent
cells that were often observed (Paper I), we might reasonably expect
that the South Pole should be almost the ideal site for the
application of relatively simple low-order image correction methods,
even despite the large amplitude of the fluctuations observed.

The concentration of the bulk of the seeing in the lower boundary
layer means that a low-order, turbulence conjugated, adaptive optics
system should be able to remove the large majority of the seeing over
very wide angles. The small free atmosphere component would remain
uncorrected in this scenario, and hence the possibility of predicting
the rare occurrences of poor (0.4--0.5$''$) free atmosphere seeing
would be very helpful (see \S\ref{sec:forecast}).

Clearly the tight limitations on the isoplanatic angle and coherence
time, $\theta$ and $\tau$ (see \S\ref{sec:spole}), mean that the ideal
of perfect image correction of turbulence-distorted images is
extremely difficult to achieve at any site, including the South
Pole. A large number of independent sub-mirrors is required to correct
the wavefront errors over the full field of view of the telescope. The
strategy at other sites has generally been to settle for varying
degrees of partial image correction, usually by ignoring the higher
orders of the perturbations of the wavefront, and thus making the
corrections applicable over a larger area. The aim is to make the
maximum gains possible in image quality with a relatively simple and
inexpensive system such as a tip-tilt mirror (i.e.\ first-order
corrections). Evidently, the maximum ``reconstruction angle'' (the
term coined by Cowie \& Songaila 1988) depends on the precision
required in the corrections, and there is a trade-off between image
quality and sky coverage.

These ideas are particularly important in relation to the situation at
the South Pole since high-altitude turbulence has a higher spatial
frequency, when viewed from ground-level, than turbulence close to the
surface. We may expect, therefore, that a low-order adaptive optics
system that corrects the boundary layer component of the seeing,
leaving the upper atmosphere uncorrected, would be effective over a
substantially greater area of the sky at the South Pole than a similar
system operating at the best mid-latitude sites.  A glance at
Fig.~\ref{ao_ir} indicates that image resolution of 0.2--0.3$''$
should be obtainable over much larger angles at the South Pole than at
Cerro Paranal.

Although no vertical $C_{\rm N}^2$ profiles were available from other
mid-latitude sites for analysis, the comparison should be much the
same as Paranal. Indeed, it may be even more favourable in comparison
with Mauna Kea, where a much greater proportion of the seeing is due
to upper atmosphere turbulence.


It is worth also mentioning some of the methods used for correction of
higher-order components of the wavefront phase fluctuation. With a
characteristic coherence length of $r_0$, the number of independent
elements required for full image correction is on the order of
$(D/r_0)^2$, for corrections made in planes conjugate to the telescope
aperture. Such a project is not expected to be any easier at the South
Pole than other sites.

A ``multiconjugate'' approach (e.g.\ Tallon et al.\ 1992) requires, in
principle, a single correction element for each turbulent layer,
which, at any site, is likely to result in a less complex system than
the usual aperture-conjugated case.  It is a particularly powerful
method in situations where the bulk of the image degradation is
produced by a small number of turbulent layers. This is clearly the
case at the South Pole, where the boundary layer seeing usually
consists of 2--4 intense layers, with a similar number, or less, of
much weaker layers in the free atmosphere. It is likely, therefore,
that the South Pole represents a particularly good site for
higher-order corrections using such a technique.

\section{Prospects for Super Seeing from the Summits of the Plateau}
In the broadest sense, the aim of the site-testing campaign underway
in Antarctica (e.g.\ Storey, Ashley \& Burton 1995) is to find the
best site for astronomy in Antarctica. Given that the only
measurements of site conditions available at this point are from the
South Pole, it is important to use the results of this experiment to
attempt to draw some conclusions about the likely seeing conditions
elsewhere on the high plateau. It is generally agreed that the South
Pole is almost certainly not the best site for astronomy in
Antarctica. Indeed, the only reason for choosing this location for the
experiments conducted so far is that it has been the only place on the
plateau that has been easily accessible. In any case, the conditions
there should provide a good indication of the magnitude of the
improvements that might be expected from the very best sites, from
infrared to millimetre wavelengths.

The South Pole lies a long way off the central ``ridge'' that marks
the highest elevations on the plateau. The implications of this are
quite obvious if we consider the three key words: ``high'', ``cold''
and ``dry''. Dome~A (82$^{\circ}$\,S, 80$^{\circ}$\,E), at 4,200\,m,
and some 1,000\,km from the South Pole, is the highest point on the
plateau. Although no data are available from Dome~A (in fact, it is
uncertain whether anyone has ever set foot there), indications from
the closest stations are that it should be at least 10$^{\circ}$C
colder than the South Pole on average. Given the enormous difference
between the South Pole and the best mid-latitude sites in the thermal
infrared (Chamberlain et al.\ 2000, Phillips et al.\ 1999, Nguyen et
al.\ 1996), we might expect further reductions again in the thermal
background from Dome~A\@. In addition, the thinner atmosphere
($\sim$5,000\,m physiological altitude at the surface) and lower
temperature should result in even lower water vapour content than the
South Pole, which is already a factor five times lower than Mauna Kea.

These ideas are widely known now. What is less certain is the quality
of the seeing we might expect higher on the plateau. Since the seeing
depends on a more complex interplay of atmospheric conditions, it is
not as easy to predict any improvements, let alone what the scale of
those improvements might be. In this section we examine the available
data from elsewhere in East Antarctica, and compare this with what we
have discovered at the South Pole, in an attempt to give some
indication of what we might expect to find at the higher elevations.

The paucity of information about atmospheric conditions over East
Antarctica make this a difficult task. Reasonably continuous,
long-term records of both surface and upper air meteorology parameters
are available from only two places: the South Pole and Vostok (of
which only the South Pole has continuously maintained a winter crew in
recent years). These have been supplemented by a few Automatic Weather
Stations (AWS), successfully deployed at other locations above 3,000~m
for periods of a few years at a time, which provide some valuable
additional information about surface conditions. Overall, we have
access to data from no more than four or five places, in an area half
the size of Australia. Many of the results are discussed in the book
of Schwerdtfeger (1984), and it is from this source that much of the
weather information used here has been
gleaned. Table~\ref{plateauweather} is a summary of all the
measurements that could be found that are in some way related to the
seeing (in particular, the boundary layer seeing). Some of the numbers
quoted are based on no more than a couple of years data, especially
from the AWSs, and so any conclusions made here on the basis of these
figures should be treated with caution.

\begin{table*}
\begin{center}
\begin{tabular}{lccccc}
\hline\noalign{\smallskip}
 & South Pole & Vostok & Plateau & Dome C & Dome A \\
\hline\noalign{\smallskip}
Years & 26 (40) & 24 & 3 & 8 & -- \\
Altitude (m) & 2835 & 3488 & 3625 & 3280 & 4200 \\
Latitude ($^{\circ}$S) & 90 & 78.5 & 79.2 & 74.5 & 82 \\
Yearly Mean Temperature ($^{\circ}$C) & $-49.3$ & $-55.4$ & $-56.4$ & $-50.6$ & $\sim -60$ \\
Coldest Monthly Mean & $-59.9$ & $-68.3$ & $-71.4$ & $-61.7$ & -- \\
\hspace*{0.5cm}Temperature ($^{\circ}$C) & (Jul) & (Aug) & (Aug) & (Aug) & -- \\
$\delta T_{\rm BL} \ (^{\circ}$C) & 20 & 23 & $\sim 20$ & $\sim 20$ & $> 25$ \\
Slope (degrees) & $1.0\times10^{-3}$ & $1.3\times10^{-3}$ & $0.8\times10^{-3}$ & $\sim 0$ & $\sim 0$ \\
$\bar{V}$ (m\,s$^{-1}$) & 5.8 & 5.1 & -- & 2.8 & -- \\
Constancy & 0.79 & 0.81 & 0.67 & 0.53 & -- \\
Calm (\%) & 2 & 1 & -- & 10 & -- \\
$V_{\rm max}$ (m\,s$^{-1}$) & 24 & 25 & -- & 16 & -- \\
Cloud $< \frac{3}{10}$ (winter) (\%) & 63 & 56 & 65 & -- & -- \\
\noalign{\smallskip}
\hline
\end{tabular}
\end{center}
\caption[ ]{Comparison of weather parameters at all sites on the 
high plateau for which information is available. Data on Dome~C are
taken from Keller et al.\ (1991, 1993, 1995), values for for Plateau
and Vostok stations are from Schwerdtfeger (1984). `Years' denotes the
length of time for which records are available, `$\delta T_{\rm BL}$'
is the temperature change across the boundary layer, `Slope' is the
mean slope of the surface, `$\bar{V}$' is the mean surface wind speed,
`Constancy' is ratio of the average wind velocity to the average wind
speed, `$V_{\rm max}$' is the maximum wind speed ever recorded, and
`Cloud' gives the percentage of time in winter where the cloud cover
is less than 30\%. Where no data is available the entry is indicated
with a ``--''.}
\label{plateauweather}
\end{table*}

The boundary layer seeing at the South Pole is generally intense, but
highly variable and critically dependent upon the interplay of the
temperature gradient and inversion wind, in particular any vertical
irregularities that occur. The $C_{\rm N}^{2}$ signal tends to be most
intense close to the surface. Above the boundary layer, the free
atmosphere is in general exceptionally calm and clear. Whilst no
microthermal data are available from any of the sites listed in
Table~\ref{plateauweather}, aside from the Pole, these meteorological
parameters may offer some clues as to the likely structure of the
boundary layer at these sites.

One factor that does differ from the South Pole to the other sites is
diurnal variation. At Plateau Station over a 24\,hour period the depth
of the temperature inversion falls from around 15$^{\circ}$ at night
to $<$ 5$^{\circ}$ during the day (Schwerdtfeger 1984). Although the
diurnal changes appear to be mostly restricted to the lowest
10--15\,m, according to the data from the South Pole, this is probably
the most turbulent region in the whole boundary layer. Given that the
winds at surface level at a site such as Plateau Station are driven by
the inversion, it is likely that these too undergo some diurnal
variation. The impact of this on the stability of the boundary layer
in terms of turbulence is not clear, but certainly, at mid-latitude
sites, it is a general rule that the best seeing is observed later in
the evenings, when the boundary layer has stabilised. Such temperature
variations are likely to have some minor effect in Antarctica,
increasing as the latitude of the site decreases. Since all of the
highest sites are south of 75$^{\circ}$~S, there are at least a few
months everywhere during the winter when the Sun doesn't rise, and
little or no diurnal effect would be noticed during this period.

The increasing strength of the inversion at higher altitudes means
that, in fact, the temperature of the warmest layer above the site is
very similar from place to place. Comparison of weather balloon data
from the South Pole and Vostok show that the temperatures at the
500\,hPa level at each station in winter differ by less than a degree,
with the South Pole being slightly warmer. The temperature inversion
at the South Pole is about 400\,m high, according to the temperature
data from our microthermal balloon sondes (in terms of $C_{\rm
N}^{2}$, it levels off at about 200--220\,m). While no information is
available regarding the vertical extent of the inversion at Vostok, it
is likely that it is slightly narrower at the higher altitude site,
given that it reaches almost the same temperature at the same pressure
altitude. If the temperature layers are to some degree stratified
across the plateau, according to pressure altitude, then it may be
that Dome~C also has a somewhat narrower inversion, while at Dome~A it
could be significantly narrower again.

Without discussing wind characteristics at each site for a moment, the
results of our $C_{\rm N}^{2}$ measurements of boundary layer
turbulence at the South Pole are clearly associated with vertical
fluctuations in d$T$/d$z$ in the boundary layer. The seeing contributions
are generally proportional to the magnitude of the temperature
fluctuations, for a given strength of wind shear. This indicates that
individual turbulent disturbances in the boundary layer at higher
sites may be greater in magnitude than those observed at the South
Pole due to the combined effect of both of these differences: a
stronger inversion, extending over a smaller vertical distance, means
that the temperature {\em gradient} may be significantly steeper,
especially close to the surface. Any mechanical turbulence, then, of a
similar sort to that experienced at the South Pole would naturally
produce more intense turbulent cells, in terms of $C_{\rm N}^{2}$.

A narrower inversion also has implications from the point of view of
image correction techniques. We now know that the concentration of
optical turbulence at the South Pole very close to the surface leads
to very favourable conditions for the application of adaptive
optics. If the boundary layer at the higher sites does turn out to
extend over a shorter distance, this argument applies even more
forcefully, regardless of the baseline seeing. Our conclusion for the
South Pole was that a system designed to correct only the boundary
layer seeing would be applicable over a characteristic angle of about
1--2$'$ in the visible. Hence, any significant reduction of the height
of the boundary layer at a place like Vostok or Dome~C could result in
a large increase in the angular scale, and hence the sky coverage, of
adaptive optics-type image correction techniques.

What is hoped for, of course, is that the natural seeing will be
better at other sites on the plateau, and the wind conditions must be
considered before this possibility can be dismissed. A conclusion that
came through strongly from our South Pole data was that fairly strong
wind shears within the boundary layer are a necessary ingredient for
strong optical turbulence. It is possible to have clear seeing through
a deep inversion if there are no strong wind shear regions; peaks in
d$T$/d$z$ will not do the damage on their own. The available wind data
from other sites on the plateau suggest that, here too, the effect on
the seeing is likely to be somewhat different to that observed at the
South Pole.

South Pole Station is located well down from the peak of the plateau,
and lies on a gentle slope with a gradient of approximately 1 part in
$10^3$. This is roughly the same as Vostok and Plateau stations.
These gentle slopes are enough to generate an inversion wind with an
average speed of just over 5\,m\,s$^{-1}$ at both the South Pole and
Vostok (see Table~\ref{plateauweather}). The directional constancy is
very high: around 0.8 at Pole and Vostok, slightly lower at Plateau,
with the vectored average wind blowing in a direction oriented at
about 30$^{\circ}$ to the fall line of the terrain. This is an
indication that the ubiquitous inversion wind dominates the surface
flow, but is modified in direction, to some extent, by the lesser
effects of synoptic air flows and the coriolis effect. The calculated
average flow lines for the surface winds across the continent (see
Fig.~3 from Paper II) illustrate the importance of the local
topography on the wind characteristics at any given location.

Since the geostrophic winds above the boundary layer do not generally
blow from the same direction as the inversion wind, at the South Pole
at least, there are inevitably shear layers in the boundary layer,
where the winds change both speed and direction. One of our main
conclusions was that virtually all of the turbulence responsible for
the seeing occurs in layers where wind shear causes disturbances in
the vertical temperature gradient. This statement describes almost all
of the observed optical turbulence at the South Pole, and explains why
the overwhelming majority of the seeing arises from the boundary
layer.

The situation at Dome~C is rather different. Here, the terrain is
almost perfectly flat, and the inversion wind is negligible. Wind
speeds are generally much lower (2.8\,m\,s$^{-1}$ on average), with a
lower constancy factor (about 0.5), reflecting the fact that the wind
direction is determined by weather patterns rather than the slope of
the terrain. A more detailed study of some of the monthly AWS data
reveals that Dome~C enjoys winds of less than 2\,m\,s$^{-1}$ roughly
50\% of the time during winter, and below 4\,m\,s$^{-1}$ up to 80\% of
the time. This compares with approximate figures of 10\% and 50\%,
respectively, at the South Pole. Dome~C is a very calm site.

The absence of an inversion wind means that this source of turbulence,
at least, can be discounted. The kinds of wind shear layers observed
at the South Pole ought not exist. There may be other sources of
vertical wind gradients, such as those observed occasionally in the
free atmosphere over the South Pole. There is, however, a possibility
that there is in fact no significant source of mechanical turbulence
in the boundary layer, in which case the seeing could be very good
from ground level. Dome~C, along with Dome~A, may be an oasis of calm
in the turbulent inversion layer stretching right across the plateau.
 
Another feature of the South Pole seeing is the remarkably quiescent
free atmosphere. Upper-level jet streams associated with the
tropopause are virtually absent in the data, and very few other
significant sources of turbulence were observed. From the long-term
records, the strong upper-troposphere winds (jet streams) that limit
the seeing at mid-latitude sites are virtually absent. At the 300\,hPa
level (about 10\,km altitude at the South Pole), winds of over
30\,m\,s$^{-1}$ occur 5\% of the time, and over 40\,m\,s$^{-1}$ a mere
1\%. This is compared with 14\%~/~3\% at Byrd Station
(80$^{\circ}$~S), and larger values further north. So it would appear
that the free atmosphere is somewhat less stable as one moves out from
the centre of the polar vortex, and there would probably be a
corresponding increase in upper-atmosphere turbulence. Hence the
average free atmosphere may be somewhat worse than the very low value
of around $0.3''$ (upward from 200\,m) measured at the South Pole.

\section{Summary}
To summarise our results, the microthermal turbulence at the South
Pole is concentrated much closer to the surface than at the best
mid-latitude sites. So, while the integrated seeing from surface level
is poor, the free atmosphere is quieter than at these other
sites. This has the result that the seeing contribution above about
200~m (the depth of the South Pole boundary layer) is significantly
less than at any temperate latitude site.

The available evidence also suggests that the high-altitude sites such
as Domes~A and C probably experience better boundary layer seeing than
the South Pole. Given that this is by far the dominant source of image
degradation at the Pole, this is a very important result. The
conclusion is based on the absence of the inversion winds that
generate much of the wind shear responsible for the mechanical
turbulence.

In addition, the inversion layers themselves may be significantly
narrower at the higher sites. The narrow turbulent region at the South
Pole, relative to mid-latitude sites, greatly increases the
characteristic angles over which image correction techniques can be
applied. The higher sites may also, therefore, be significantly better
in terms of their potential for adaptive optics.

\begin{acknowledgements}
This paper is an edited version of unpublished results from Rodney
Marks' PhD thesis, prepared by Michael Burton, with assistance from
Michael Ashley and John Storey from the University of New South Wales
and Darryn Schneider from the Antarctic Muon and Neutrino Detector
Array (AMANDA) project.  This research was supported by an Australian
Postgraduate Award and by grants from the Australian Research Council
and the Australian Department of Industry, Science and Technology.
Rodney worked closely during his PhD with Jean Vernin and others from
the University of Nice, Bob Pernic and colleagues from the US Center
for Astrophysical Research in Antarctica, and with many colleagues
from the University of New South Wales.  Astronomy is much richer for
the spirit of international cooperation that Rodney fostered through
these collaborations.

\end{acknowledgements}

\end{document}